\documentclass{vldb}

  \usepackage{times,amsmath,epsfig,subfigure,stmaryrd}

   \usepackage{tikz}
   \usetikzlibrary{matrix,arrows}
   \usepackage{multirow}
   \usepackage{colortbl}
   \usepackage{color}
   \usepackage{listings}   

\newcounter{enum}
\newenvironment{packed_enum}{
\begin{list}{\textbf{(\arabic{enum}.)}}{
  \setlength{\itemsep}{0pt}
  \setlength{\parskip}{0pt}
  \setlength{\labelwidth}{-4 pt}
  \setlength{\leftmargin}{0 pt}
  \setlength{\itemindent}{0pt}
  \usecounter{enum}}
}{\end{list}}

\newcommand{\no}[1]{
\definecolor{shadecolor}{rgb}{0.32549019607843,0.32549019607843,0.32549019607843}
\hspace*{-1.9ex}
\raisebox{-0.5ex}[0cm][0cm]{
\hspace*{-.9ex}
\begin{tikzpicture} [scale=1,every node/.style={circle,draw=black!70,text=black,thick,font=\Large,scale=.45}]
\node at (0,0) {#1} ; \\
\end{tikzpicture}
\hspace*{-1.2ex}}
}

\definecolor{orange}{rgb}{1,0.5,0}
\definecolor{lightpurple}{rgb}{0.5,0,0.5}
\definecolor{purple}{rgb}{0.5,0,0.25}
\definecolor{mybrown}{rgb}{0.5,0,0}
\definecolor{lightgray}{rgb}{0.9,0.9,0.9}
\definecolor{lightgreen}{rgb}{0.7,0.9,0.3}

\definecolor{revisioncolor}{rgb}{0,0,1}
\definecolor{crcolor}{rgb}{0,1,0}
\definecolor{cr2color}{rgb}{0.58,0,0.82}
\definecolor{dropcolor}{rgb}{1,0,0}

\newcommand{\hide}[1]{}

\newcommand{\hail}{LIAH}

\usepackage[ruled,vlined]{algorithm2e}

\usepackage{graphicx}
\usepackage[pdftex,colorlinks=true,citecolor=black,urlcolor=black,linkcolor=black]{hyperref}

\begin{document}

\title{Survival of the Fastest: Adaptive Indexing in MapReduce}
\title{Only Lazy Elephants are Robust Elephants}
\title{Aggressive Elephants Also Have to Adapt to Survive}
\title{Survival of the Fastest: \\Only Fast Adaptive Elephants Survive}
\title{Even Aggressive Elephants Must Adapt}
\title{Even Aggressive Elephants Must Adapt to Stay Fast}
\title{LIAH: Lazy Indexing and Adaptivity in Hadoop}
\title{Towards Zero-Overhead Adaptive Indexing in Hadoop}
\numberofauthors{1}
\author{
\alignauthor Stefan Richter, Jorge-Arnulfo Quian\'e-Ruiz, Stefan Schuh, Jens Dittrich \\
\vspace{0.2cm}
       \affaddr{Information Systems Group\\Saarland University}\\
       \affaddr{http://infosys.cs.uni-saarland.de}\\
}
\maketitle

\begin{sloppypar}

\begin{abstract}
Several research works have focused on supporting index access in MapReduce systems. These works have allowed users to significantly speed up selective MapReduce jobs by orders of magnitude. However, all these proposals require users to create indexes upfront, which might be a difficult task in certain applications (such as in scientific and social applications) where workloads are evolving or hard to predict. To overcome this problem, we propose \hail\ ({\it Lazy Indexing and Adaptivity in Hadoop}), a parallel, adaptive approach for indexing at minimal costs for MapReduce systems. The main idea of \hail\ is to automatically and incrementally adapt to users' workloads by creating clustered indexes on HDFS data blocks as a byproduct of executing MapReduce jobs. Besides distributing indexing efforts over multiple computing nodes, \hail\ also parallelises indexing with both map tasks computation and disk I/O. All this without any additional data copy in main memory and with minimal synchronisation. The beauty of \hail\ is that it piggybacks index creation on map tasks, which read relevant data from disk to main memory anyways. Hence, \hail\ does not introduce any additional read I/O-costs and exploit free CPU cycles. As a result and in contrast to existing adaptive indexing works, \hail\ has a very low (or invisible) indexing overhead, usually for the very first job. Still, \hail\ can quickly converge to a complete index, i.e.~all HDFS data blocks are indexed. Especially, \hail\ can trade early job runtime improvements with fast complete index convergence. We compare \hail\ with HAIL, a state-of-the-art indexing technique, as well as with standard Hadoop with respect to indexing overhead and workload performance. In terms of indexing overhead, \hail\ can completely index a dataset as a byproduct of only four MapReduce jobs while incurring a low overhead of $11\%$ over HAIL for the very first MapReduce job only. In terms of workload performance, our results show that \hail\ outperforms Hadoop by up to a factor of $52$ and HAIL by up to a factor of $24$.
\end{abstract}

\section{Introduction}
\label{introduction}

In recent years, a huge number of research works have focused on improving the performance of Hadoop MapReduce~\cite{cheetah,hadooppp,ifAnalysisMR,optimizationMR,trojanLayouts}. In particular, several researchers have focused on supporting efficient index access in Hadoop~\cite{traverse,fulltextindex,mapredPerf}. Some of these works have improved the performance of selective MapReduce jobs by orders of magnitude. However, all these indexing approaches have three main weaknesses. First, they require a high upfront cost or long idle times for index creation. Second, they can support only one physical sort order (and hence one clustered index) per dataset. Third, they require users to have a good knowledge of the workload in order to choose the indexes to create. 

Recently, we proposed HAIL~\cite{hail} ({\it Hadoop Aggressive Indexing Library}) to solve the first two problems, i.e.~high upfront indexing costs and lack of supporting multiple sort orders. HAIL allows users to create multiple clustered indexes at upload time almost for free. As a result, users can speed up their MapReduce jobs by almost two orders of magnitude. But, this improvement only happens if users create the right indexes when uploading their datasets to HDFS. This means that, like traditional indexing techniques~\cite{physicaldb,costdriving,dbtuning,physicaldesign,selftuning,traverse,fulltextindex,hadooppp,mapredPerf}, HAIL requires users to decide upfront which indexes to create. Thus, HAIL as well as traditional indexing techniques are not suitable for unpredictable or ever-evolving workloads~\cite{mrtutorial}. In such scenarios, users often do not know which indexes to create beforehand. Scientific applications and social networks are a clear example of such use-cases~\cite{nodb}.

\subsection{Motivation}
\label{introduction_motivation}
Let us see through the eyes of a group of scientists, say Alice and her colleagues, who want to analyse their daily experimental results using Hadoop MapReduce. Basically, the experimental results are collected in a large dataset (typically in the order of terabytes) containing many dozens of numeric attributes. To understand and interpret the experimental results, Alice and her colleagues navigate through the dataset according to the properties and correlations of the data~\cite{nodb}. The problem is that Alice and her colleagues typically: (i)~do not know the data access patterns in advance; (ii)~have different interests and hence cannot agree upon common selection criteria at data upload time; (iii)~even if they agree which attributes to index at data upload time, they might end up filtering records according to values on different attributes. Therefore, HAIL (as well as traditional indexing techniques) cannot help Alice and her colleagues, because HAIL is still a static system that cannot adapt to changes in query workloads.

One day Alice hears about {\it adaptive indexing}~\cite{dbcracking}, where the general idea is to create indexes as a side-effect of query processing. Adaptive indexing aims at creating indexes incrementally in order to avoid high upfront index creation times. Alice is excited about the adaptive indexing idea since this could solve her (and her colleagues') problem. However, Alice notices that she cannot apply existing adaptive indexing works~\cite{DFK05,dbcracking,crakingself,adaptivemerging,mergecracking,adaptivesthocastic} in MapReduce systems for several reasons:

First, these techniques aim at converging to a global index for an entire attribute, which requires sorting the attribute globally. Therefore, these techniques perform many data movements across the entire dataset. Doing this in MapReduce would hurt fault-tolerance as well as the performance of MapReduce jobs. This is because we would have to move data across HDFS data blocks\footnote{Henceforth, we refer to an HDFS data block simply as data block.} in sync with all their three physical data block replicas.

Second, even if Alice applied existing adaptive indexing techniques inside data blocks, these techniques would end up in many costly I/O operations to move data on disk. This is because these techniques consider main-memory systems and thus do not factor in the I/O-cost for reading/writing data from/to disk. Only one of these works~\cite{adaptivemerging} proposes an adaptive merging technique for disk-based systems. However, applying this technique inside a data block would not make sense in MapReduce since data blocks are typically loaded entirely into main memory anyways when processing map tasks. One may think about applying adaptive merging across data blocks, but this would again hurt fault-tolerance and the performance of MapReduce jobs as described above.

Third, these works focus on creating unclustered indexes in the first place and hence it is only beneficial for highly selective queries. One of these works~\cite{crakingself} introduced lazy tuple reorganisation in order to converge to clustered indexes. However, this technique needs several thousand queries to converge and its application in a disk-based system would again introduce a huge number of expensive I/O operations.

Fourth, existing adaptive indexing approaches were mainly designed for single-node DBMSs. Therefore, applying these works in a distributed parallel systems, like Hadoop MapReduce, would not fully exploit the existing parallelism to distribute the indexing effort across several computing nodes.

\subsection{Idea}
\label{introduction_adaptive}
We propose {\it \hail} ({\it Lazy Indexing and Adaptivity in Hadoop}): a lazy and adaptive indexing approach for parallel disk-based systems, such as MapReduce. The main idea behind \hail\ is to exploit the existing MapReduce pipeline in order to build clustered indexes in a scalable, automatic, and almost invisible way as byproduct of job executions. For this, \hail\ interprets incoming jobs as hints about what might be a worthwhile index. A salient feature of \hail\ is that it piggybacks on job execution in such a way that no additional read I/O-cost is required for indexing purposes. In other words, \hail\ only requires some additional I/O-cost for writing clustered indexes back to disk. This allows \hail\ to quickly converge to a {\it complete index}, i.e.~all HDFS data blocks are indexed, with a very low indexing overhead. Another interesting feature is that \hail\ stores a clustered index created at query processing time in an additional HDFS file, called {\it pseudo data block replica}. In fact, a pseudo data block replica is another logically indexed replica for a given data block. Therefore, pseudo data block replicas allow us to support a different number of replicas for each data block, which is crucial for incremental indexing.

Like existing adaptive indexing works, \hail\ distributes the indexing effort over several queries to avoid negatively impacting the performance of an individual query. However, \hail\ differs from existing adaptive indexing works in four major aspects:

First, \hail\ focuses on block level clustered indexes. This means that \hail\ creates one clustered index for each data block instead of a single index per attribute. Consequently, \hail\ reorders data only inside a data block, which preserves the fault-tolerance of Hadoop since data is never shuffle across data blocks.
Second, \hail\ parallelises the adaptive indexing effort across several computing nodes in order to limit the indexing overhead. 
Third, \hail\ considers disk-based systems and hence it factors in the cost of reading from and writing to disk. \hail\ completely sorts a data block once it reads the block from disk. This avoids future expensive I/O operations for refining the index inside a data block. Still, \hail\ does not sort all data blocks in one pass in order to avoid that a single MapReduce job pays the high cost of index creation. Notice that adaptive merging~\cite{adaptivemerging}, which also considers disk based systems, is orthogonal to the focus of \hail. While \hail\ produces a set of sorted partitions incrementally, adaptive merging aims at incrementally combining such sorted partitions. Fourth, \hail\ creates clustered indexes rather than unclustered indexes. This allows \hail\ to benefit from index scans even for lowly selective jobs. Since \hail\ stores datasets in PAX representation~\cite{pax}, creating clustered indexes also allows LIAH to avoid expensive random read I/O operations for tuple reconstruction.

\subsection{Research Challenges}
\label{introduction_challenges}

The approach followed by \hail\ triggers a number of interesting research challenges:
\begin{packed_enum}
\item How can we change the job execution pipeline to create clustered indexes at job execution time? How to index big data incrementally in a disk-based system? How to minimise the impact of indexing on job execution times? How to efficiently interleave data processing with indexing? How to create several clustered indexes for read-only data blocks at query time? How to support different number of replicas per data block? How will the job execution pipeline change for Alice and her colleagues?

\item How can we change Hadoop to exploit newly created clustered indexes? How to distribute the indexing effort efficiently by considering data-locality and index placement across computing nodes? How to schedule map tasks to efficiently process indexed and non-indexed data blocks without affecting failover? How will jobs change from the perspective of Alice and her colleagues?
\end{packed_enum}

\subsection{Contributions}
\label{introduction_contributions}
We present \hail, a lazy and adaptive indexing approach for MapReduce systems. The main goal of \hail\ is to minimise the impact of indexing on job execution times. We make the following four contributions:
\vspace{-.2cm}
\begin{packed_enum}
\item We show how to effectively piggyback adaptive index creation on the existing MapReduce job execution pipeline. In particular, we show how to parallelise indexing with both the computation of map tasks and disk I/O. All this without any additional data copy in main memory and minimal synchronisation. A particularity of our approach is that we always index a data block entirely, i.e.~in a single pass. As a result, \hail\ not only allows map tasks of future jobs to perform an index access, but it also frees them from costly extra I/O operations for refining indexes. 

\item We show how to efficiently process pseudo data block replicas, i.e.~data block replicas containing a clustered index adaptively created by \hail. The beauty of our approach is that it is completely invisible from the users' perspective. \hail\ takes care of performing MapReduce jobs using normal data block replicas or pseudo data block replicas (or even both).
Additionally, \hail\ comes with its own scheduling policy, called \hail\ Scheduling. 
The idea of \hail\ Scheduling is to balance the indexing effort across computing nodes so as to limit the impact of indexing on job runtime. As a side effect of balancing the indexing effort, \hail\ improves parallel index access for future jobs as indexes are balanced across nodes.


\item We propose a set of indexing strategies that makes \hail\ aware of the performance and the selectivity of MapReduce jobs. We first present {\it eager adaptive indexing}, a technique that allows \hail\ to quickly adapt to changes in users' workloads at a low indexing overhead. In particular, eager adaptive indexing allows \hail\ to trade early job runtime improvements with fast complete index convergence. Next, we show how \hail\ can decide which data blocks to index based on the selectivities of jobs. Then, we present the {\it invisible projection} technique that allows \hail\ to efficiently create clustered indexes for jobs having different attribute projections. Additionally, in Appendix~\ref{appendix:lazy_reordering}, we present a {\it lazy projection} technique that allows \hail\ to integrate an attribute into a clustered index only when the attribute is accessed by a job.

\item We present an extensive experimental comparison of \hail\ with Hadoop and HAIL~\cite{hail}. We use two clusters, each having different types of CPU. A series of experiments shows the superiority of \hail\ over both Hadoop and HAIL. In particular, our experimental results demonstrate that \hail\ quickly adapts to query workloads with a negligible indexing overhead. Our results also show that \hail\ has a low overhead over Hadoop and HAIL for the very first job only: all the following jobs are faster in \hail.
\end{packed_enum}

\section{Related Work}
\label{relatedwork}

\noindent{\bf Offline Indexing.}
Indexing is a crucial step in all major DBMSs~\cite{physicaldb,costdriving,dbtuning,physicaldesign,selftuning}. The overall idea behind all these approaches is to analyze a query workload and decide which attributes to index based on these observations. Several research works have focused on supporting index access in MapReduce workflows~\cite{traverse,fulltextindex,hadooppp,mapredPerf}. However, all these offline approaches have three big disadvantages. First, they incur a high upfront indexing cost that several applications cannot afford (such as scientific applications). Second, they only create a single clustered index per dataset, which is not suitable for query workloads having selection predicates on different attributes. Third, they cannot adapt to changes in query workloads without the intervention of a DBA. Recently, we proposed HAIL~\cite{hail} to solve the first two problems, but HAIL is an enhancement of the upload pipeline in HDFS. Therefore, HAIL still cannot adapt to changes in the query workload. \hail\ completes the puzzle: it enhances the Hadoop MapReduce framework (and not HDFS) in order to allow Hadoop to adapt to query workloads.

\vspace{0.1cm}
\noindent{\bf Online Indexing.}
Tuning a database at upload time has become harder as query workloads become more dynamic and complex. Thus, different DBMSs started to use online tuning tools to attack the problem of dynamic workloads~\cite{colt,totune,onlineIdx,softindexes}. The idea is to continuously monitor the performance of the system and create (or drop) indexes as soon as it is considered beneficial. Manimal~\cite{manimal,optimizationMR} can be used as an online indexing approach for automatically optimizing MapReduce jobs. The idea of Manimal is to generate a MapReduce job for index creation as soon as an incoming MapReduce job has a selection predicate on an unindexed attribute. Online indexing can then adapt to query workloads. However, online indexing techniques require to index a dataset completely in one pass. Therefore, online indexing techniques simply transfer the high cost of index creation from upload time to query processing time.

\vspace{0.1cm}
\noindent{\bf Adaptive Indexing.}
\hail\ is inspired by database cracking~\cite{dbcracking}, which aims at removing the high upfront cost barrier of index creation. The main idea of database cracking is to start organising a given attribute (i.e.~to create an adaptive index on an attribute) when it receives for the first time a query with a selection predicate on that attribute. Thus, future incoming queries having predicates on the same attribute continue refining the adaptive index as long as finer granularity of key ranges is advantageous. Key ranges in an adaptive index are disjoint, where keys in each key range are unsorted. Basically, adaptive indexing performs for each query one step of {\it quicksort} using the selection predicates as pivot for partitioning attributes. \hail\ differs from adaptive indexing in four aspects. First, \hail\ creates a clustered index for each data block and hence avoids any data shuffling across data blocks. This allows \hail\ to preserve Hadoop fault-tolerance. Second, \hail\ considers disk-based systems and thus it factors in the cost of reorganising data inside data blocks. Third, \hail\ parallelises the indexing effort across several computing nodes to minimise the indexing overhead. Fourth, \hail\ focuses on creating clustered indexes instead of unclustered indexes. A follow-up work~\cite{crakingself} focuses on lazily aligning attributes to converge into a clustered index after a certain number of queries. However, it considers a main memory system and hence does not factor in the I/O-cost for moving data many times on disk.

\vspace{0.1cm}
\noindent{\bf Adaptive Merging.}
Another related work to \hail\ is the adaptive merging~\cite{adaptivemerging}. This approach uses standard B-trees to persist intermediate results during an external sort. Then, it only merges those key ranges that are relevant to queries. In other words, adaptive merging incrementally performs external sort steps as a side effect of query processing. However, this approach cannot be applied directly for MapReduce workflows for three reasons. First, like adaptive indexing, this approach creates unclustered indexes. Second, merging data in MapReduce destroys Hadoop fault-tolerance and hurts the performance of MapReduce jobs. This is because adaptive merging would require us to merge data from several data blocks into one. Notice that, merging data inside a data block would not make sense as a data block is typically loaded entirely into main memory by map tasks anyways. Third, it has an expensive initial step to create the first sorted runs. Recently, a follow-up work uses adaptive indexing to reduce the cost of the initial step of adaptive merging in main memory~\cite{mergecracking}. However, it considers main memory systems and still has the first two problems.

To our knowledge, this work is the first research effort to propose an adaptive indexing solution suitable for MapReduce systems.

\section{HAIL Recap}



Recall that the main goal of \hail\ is to keep the impact of index creation on job runtime minimal. This is similar to the idea of HAIL, which shows that indexing during HDFS upload is possible with basically no overhead. \hail\ inherits this feature from HAIL and exploits this feature for creating indexes at query processing time. Hence, in the following, we briefly explain the data upload and MapReduce job execution pipeline in HAIL. For details about HAIL or the Hadoop execution plan see~\cite{hail} and~\cite{hadooppp}, respectively.

\subsection{Data Upload in HAIL}
Like in Hadoop, in HAIL, the first step for a user is to upload her dataset to HDFS. During the upload process, HAIL splits datasets (i.e. files) into data blocks (usually in the size of 64MB -- 256MB). Then, for each data block, HAIL stores several replicas (three by default) on different nodes for fault tolerance and load balancing reasons. HAIL differs from Hadoop in two major aspects: 
\begin{packed_enum}
\item HAIL supports logical data block replication. In other words, HAIL can store the physical data block replicas for a given logical data block in different physical layouts as long as they contain logically the same data. This follows the same principle as Trojan Layouts~\cite{trojanLayouts}. Notice that this is in contrast to Hadoop, which considers physical data block replication (i.e.~all physical data block replicas of the same logical data block are byte-identical).

\item HAIL can create as many clustered indexes as data block replicas. This is possible as HAIL uses a logical data block replication and thus HAIL can exploit different sort orders for each physical data block replica. As a result, users can configure the clustered indexes to create for their datasets. When uploading a dataset, HAIL transforms the dataset from textual row into binary PAX~\cite{pax} representation and creates the clustered indexes as specified by users. Notice that, HAIL piggybacks index creation on the natural HDFS process of copying the data from disk to main memory. Since this process is I/O-bound, HAIL can exploit unused CPU cycles to generate the  requested indexes with basically unnoticeable overhead. HAIL keeps detailed information about created indexes as index header of data block replicas. In particular, a HAIL data block contains: (i)~a block header (containing the data length and attribute offsets), (ii)~an index header, (iii)~the index data, and (iv)~the data content. Additionally, HAIL establishes a mapping {\it \{block\_id} $\rightarrow$ {\it list$<$block\_replica\_info$>$\!\}} on the HDFS NameNode for query processing purposes. Notice that, the {\it block\_replica\_info} contains the node storing the replica and the available indexes for that replica.
\end{packed_enum}

\subsection{Job Execution in HAIL}
\label{hail_job_execution}
Executing MapReduce jobs in HAIL differs from Hadoop in three aspects:
\begin{packed_enum}
\item Users can annotate their map functions with selections and projections in order to benefit from clustered indexes and the PAX representation of data blocks.

\item In the splitting phase ({\it HailSplitting}), HAIL queries the HDFS NameNode to find out if there exists a matching index with respect to annotated selections of incoming MapReduce jobs. If a suitable index exists, the HailSplitting policy forms input splits\footnote{An input split is the data unit processed by a map task.} ({\it HAIL InputSplits}) with several data block replicas that are stored on the same node. This allows HAIL to schedule a single map task to one node and avoid the high overhead for initialising and finalising map tasks. To not impact failover, one can limit the number of data blocks in a single HAIL InputSplit based on selectivities. The idea is that processing an input split should not take longer than performing a full scan over a single data block. If no suitable index exists, HAIL then falls back to Hadoop splitting by scheduling one map task per data block.

\item A map task processes its HAIL InputSplit by reading only the projected attributes. If a suitable index exists, the map task performs an index scan over its input split. Thus, HAIL usually dramatically reduces the I/O-cost for selective jobs. If no suitable index exists, HAIL falls back to full scan, but it still benefits from the PAX layout by reading only the required attributes.
\end{packed_enum}

It is worth noting that HAIL cannot benefit from its indexes if the selection predicate is on any unindexed attribute. This is because there is no way for HAIL to adapt to query workloads. \hail\ overcomes this problem: it creates additional indexes at job runtime based on the selection predicates of incoming jobs.

\section{\hail}
In this section, we discuss the fundamentals of \hail: an approach to efficiently support {\it Lazy Indexing and Adaptivity in Hadoop}. The core idea of \hail\ is to create missing but promising indexes as byproducts of full scans in the map phase of MapReduce jobs. Our concept is to piggyback on a procedure that is naturally reading the relevant data from disk to main memory anyways. This allows \hail\ to completely save the data read cost for adaptive index creation. Another beauty of \hail\ is that it builds clustered indexes in parallel to the execution of map tasks. As map tasks are usually I/O dominated, \hail\ can then exploit under-utilised CPU cycles for index creation.

In the following, we first give a general overview of the \hail\ pipeline in Section~\ref{section:adaptivehail_pipeline}. Then, in Section~\ref{section:adaptive-indexer}, we focus on the internal components for building and storing clustered indexes. In Section~\ref{accessing_blocks}, we present how \hail\ accesses the indexes created at job runtime in a way that is transparent to the MapReduce job execution pipeline. Next, in Section~\ref{incremental_indexing}, we discuss how to make the indexing overhead over MapReduce jobs almost invisible to users. Finally, in Section~\ref{section:hail_scheduling}, we present \hail\ Scheduling, which allows us to balance index creation effort across all available nodes.


\subsection{Job Execution Pipeline}
\label{section:adaptivehail_pipeline}
Let us explain the job execution pipeline in \hail\ with an example. Assume Alice wants to analyse the experimental results of a set of experiments she ran. Recall that she collects the experimental results in a large dataset containing many numeric attributes. Assume the fortunate case that Alice studies her dataset a little bit. She identifies attributes $a$, $b$, and $c$ as the most interesting search criteria for her MapReduce jobs. As HDFS uses a default replication factor of three, Alice decides to configure \hail\ to create a clustered index on each of these three attributes at upload time. This is possible because \hail\ inherits the index creation at upload time feature from HAIL. Thus, as long as Alice sends jobs with selection predicates on $a$, $b$, or $c$, \hail\ can benefit from clustered index scans. In these cases, \hail\ behaves exactly as HAIL, i.e.~map tasks perform an index scan in order to fetch only the qualifying records from disk. However, as soon as Alice (or one of her colleagues) sends a new job (say $job_d$) with a selection predicate on a different attribute (e.g.~on attribute $d$.), \hail\ cannot benefit from index scans anymore. In contrast to HAIL, \hail\ takes this missed chances of index scans as hints on how to improve the repertoire of indexes for future jobs. \hail\ piggybacks the creation of a clustered index over attribute $d$ on the execution of $job_d$. Without any loss of generality, we assume that $job_d$ projects all attributes from its input dataset. We will drop this assumption in Section~\ref{section:invisible_projection}.

\begin{figure}[t]
\centering\includegraphics[scale=.4]{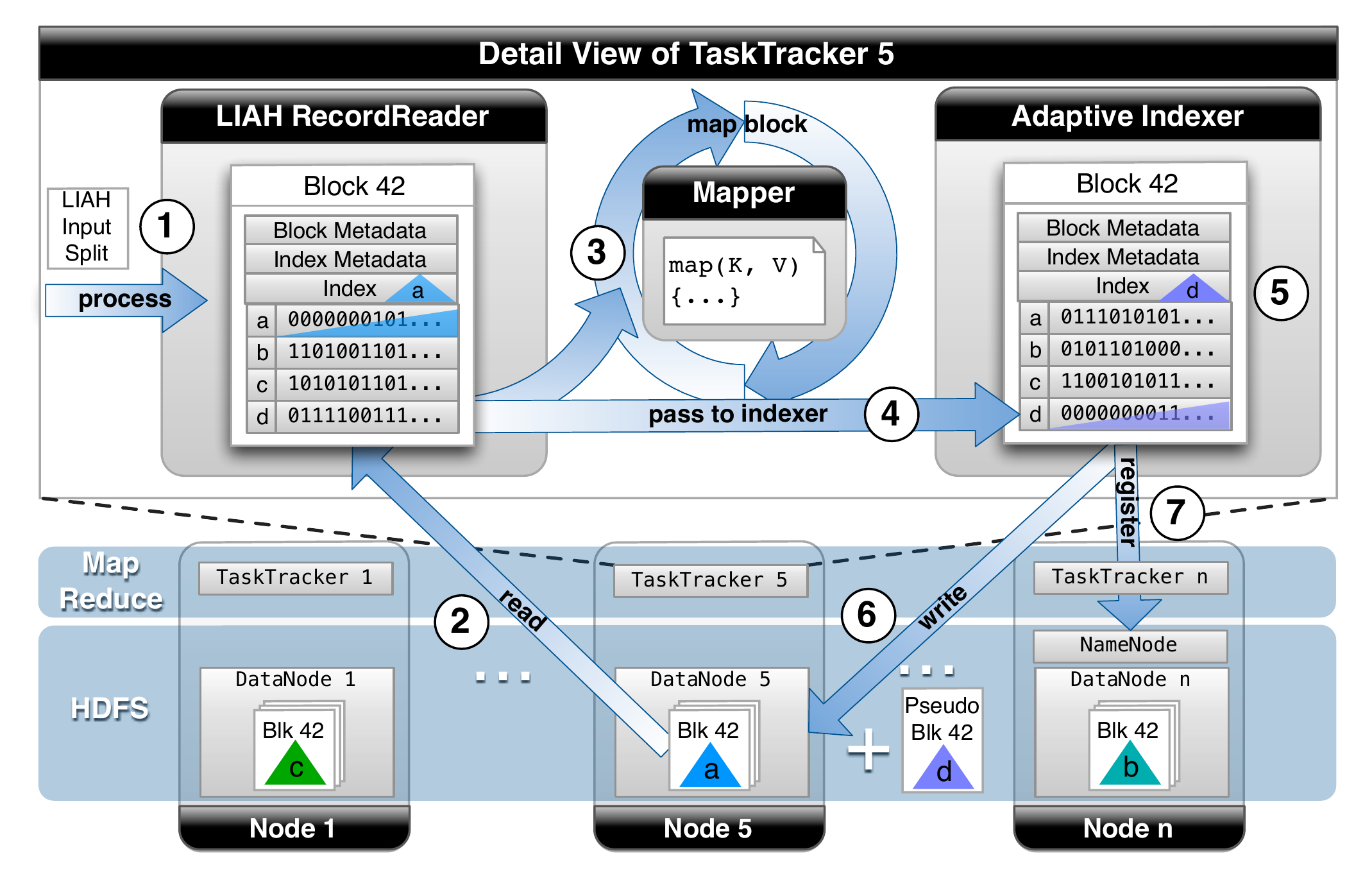}
\vspace{-0.7cm}
\caption{\hail\ pipeline.}
\label{figure:pipeline}
\vspace{-0.4cm}
\end{figure}

Figure~\ref{figure:pipeline} illustrates the general workflow of how \hail\ processes map tasks of $job_d$ when no suitable index is available. As soon as \hail\ schedules a map task to a specific TaskTracker\footnote{A Hadoop instance responsible to execute map and reduce tasks.}, e.g.~TaskTracker~5, the \hail\ RecordReader of the map task first reads the metadata (including HDFS paths, offsets, and index availability) from the \hail\ InputSplit~\no{1}. With this metadata, the \hail\ RecordReader checks whether a suitable index is available for its input data block (say $block_{42}$). As no index on attribute $d$ is available, the \hail\ RecordReader opens an input stream to the local replica of $block_{42}$ stored on DataNode~5. Then, the \hail\ RecordReader reads the metadata from the data block header to obtain the offsets of the attributes required by $job_d$. Next, the \hail\ RecordReader: (i)~loads all the values of the required attributes from disk to main memory~\no{2}; (ii)~reconstructs the records (as data blocks are in PAX representation); (iii)~feeds the map function with each record~\no{3}. Here lies the beauty of \hail: a data block that is a potential candidate for indexing was completely transferred to main memory as a natural part of the job execution process. In addition to feeding the entire $block_{42}$ to the map function, \hail\ can create a clustered index on attribute $d$ to speed up future jobs. For this, the \hail\ RecordReader passes $block_{42}$ to the {\it Adaptive Indexer} as soon as the map function finishes processing the data block~\no{4}.\footnote{Notice that, all map tasks (even from different MapReduce jobs) running on the same node interact with the same Adaptive Indexer instance. Hence, the Adaptive Indexer can end up by indexing data blocks from different MapReduce jobs at the same time.} The Adaptive Indexer, in turn, sorts the data in $block_{42}$ according to attribute $d$, aligns other attributes through reordering, and creates a sparse clustered index as described in~\cite{hail}~\no{5}. Finally, the Adaptive Indexer stores this index with a copy of $block_{42}$ (sorted on attribute $d$) as a {\it pseudo data block replica}~\no{6}. Additionally, the Adaptive Indexer registers the new created index for $block_{42}$ with the HDFS NameNode~\no{7}. In fact, the implementation of the \hail\ pipeline involves some interesting technical challenges. We discuss the \hail\ pipeline in more detail in the remainder of this section.

\subsection{Adaptive Indexer}
\label{section:adaptive-indexer}
Since \hail\ is an automatic process that is not explicitly requested by users, \hail\ should not impose unexpectedly significant performance penalties on users' jobs. Piggybacking adaptive indexing on map tasks allows us to completely save the read I/O-cost. However, the indexing effort is shifted to query time. As a result, any additional time involved in indexing will potentially add to the total runtime of MapReduce jobs. Therefore, the first concern of \hail\ is: {\it how to make adaptive index creation efficient?}

\begin{figure}[t!]
\hspace*{-0.3cm}\centering\includegraphics[scale=.4]{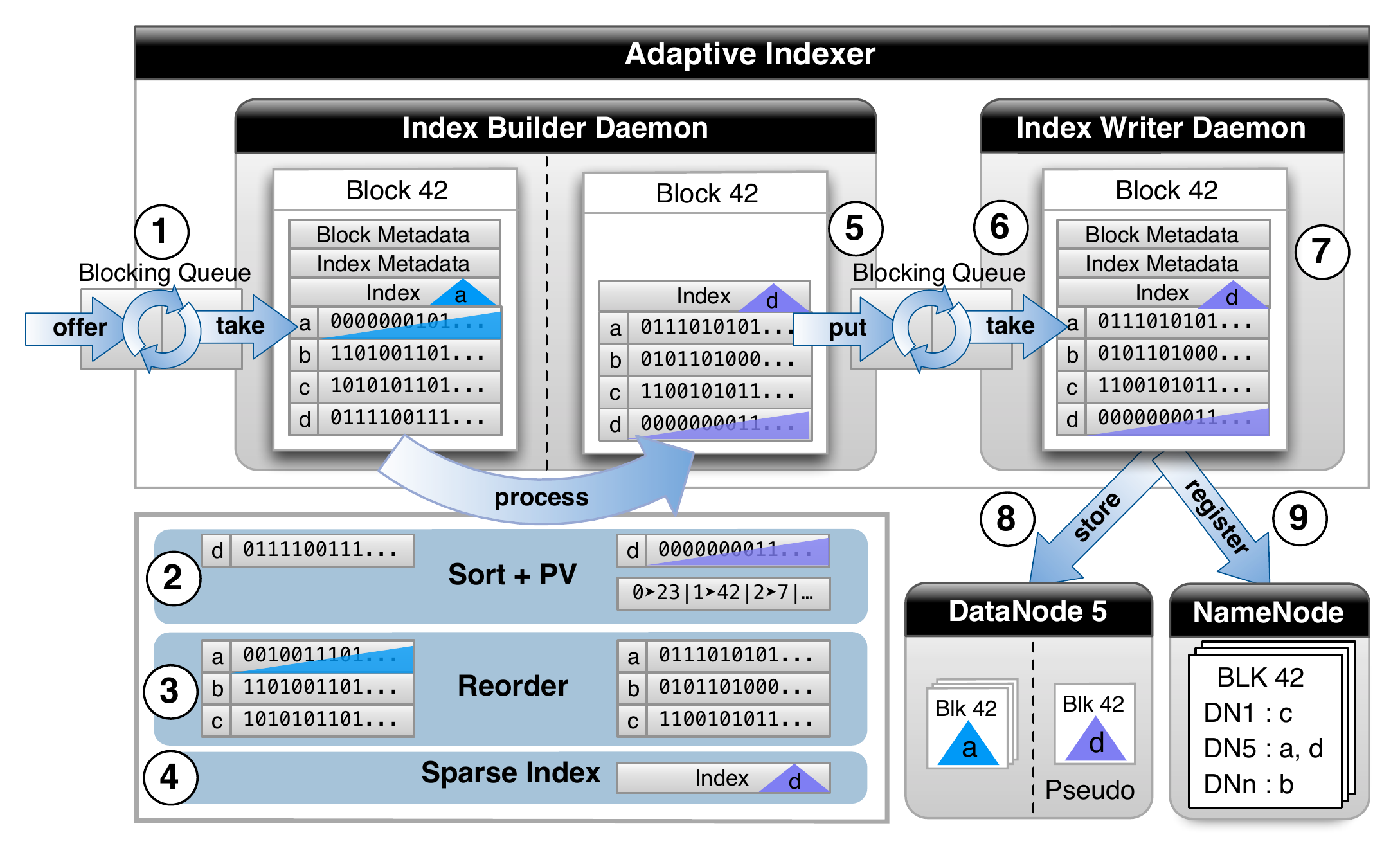}
\vspace{-0.7cm}
\caption{Adaptive Indexer internals.}
\label{figure:adaptiveindexer}
\vspace{-0.4cm}
\end{figure}

To overcome this issue, the idea of \hail\ is to run the mapping and indexing processes in parallel. However, interleaving map task execution with indexing bears the risk of race conditions between map tasks and the Adaptive Indexer on the data block. In other words, the Adaptive Indexer might potentially reorder data inside a data block, while the map task is still concurrently reading the data block. One might think about copying data blocks before indexing to deal with this issue. Nevertheless, this would entail the additional runtime and memory overhead of copying such memory chunks. For this reason, \hail\ does not interleave the mapping and indexing processes on the same data block. Instead, \hail\ interleaves the indexing of a given data block (e.g.~$block_{42}$) with the mapping phase of the succeeding data block (e.g.~$block_{43}$). For this, \hail\ uses a {\it producer-consumer} pattern: a map task acts as producer by offering a data block to the Adaptive Indexer, via a bounded blocking queue, as soon as it finishes processing the data block; in turn, the Adaptive Indexer is constantly consuming data blocks from this queue. As a result, \hail\ can perfectly interleave map tasks with indexing, except for the first and last data block to process in each node. It is worth noting that the queue exposed by the Adaptive Indexer is allowed to reject data blocks in case a certain limit of enqueued data blocks is exceeded. This prevents the Adaptive Indexer to run out of memory because of overload. Still, future MapReduce jobs with a selection predicate on the same attribute (i.e.~on attribute $d$) can at their turn take care of indexing the rejected data blocks. Once the Adaptive Indexer pulls a data block from its queue, it processes the data block using two internal components: the {\it Index Builder} and the {\it Index Writer}. Figure~\ref{figure:adaptiveindexer} illustrates the pipeline of these two internal components, which we discuss in the following.


\subsubsection{Index Builder}
\label{section:adaptive-indexer_indexbuilder}

The Index Builder is a daemon thread that is responsible for creating sparse clustered indexes on data blocks in the data queue. With this aim, the Index Builder is constantly pulling one data block after another from the data block queue~\no{1}. Then, for each data block, the Index Builder starts with sorting the attribute column to index (attribute $d$ in our example)~\no{2}. Additionally, the Index Builder builds a mapping $\{old\_position \rightarrow new\_position\}$ for all values as a permutation vector. After that, the Index Builder uses the permutation vector to reorder all other attributes in the offered data block~\no{3}. Once the Index Builder finishes sorting the entire data block on attribute $d$, it builds a sparse clustered index on attribute $d$~\no{4}. Then, the Index Builder passes the newly indexed data block to the Index Writer~\no{5}. The Index Builder also communicates with the Index Writer via a blocking queue. This allows \hail\ to also parallelise indexing with the I/O process for storing newly indexed data blocks.

\subsubsection{Index Writer}
\label{section:adaptive-indexer_indexwriter}

The Index Writer is a daemon thread that is responsible for persisting indexes created by the Index Builder to disk. The Index Writer continuously pulls newly indexed data blocks from its queue in order to persist them on HDFS~\no{6}. Once the Index Writer pulls a newly indexed data block (say $block_{42}$), it creates the block metadata and index metadata for $block_{42}$~\no{7}. Notice that a newly indexed data block is just another replica of the logical data block, but with a different sort order. For instance, in our example of Section~\ref{section:adaptivehail_pipeline}, creating an index on attribute $d$ for $block_{42}$ leads to having four data block replicas for $block_{42}$: one replica for each of the first four attributes. Therefore, the Index Writer could simply write a new indexed data block as another replica. However, HDFS supports data block replication only at the file level, i.e.~HDFS replicates all the data blocks of a given dataset the same number of times. This goes against the incremental nature of \hail.

\begin{figure*}[t]
\vspace{-0.2cm}
\centering\includegraphics[scale=.5]{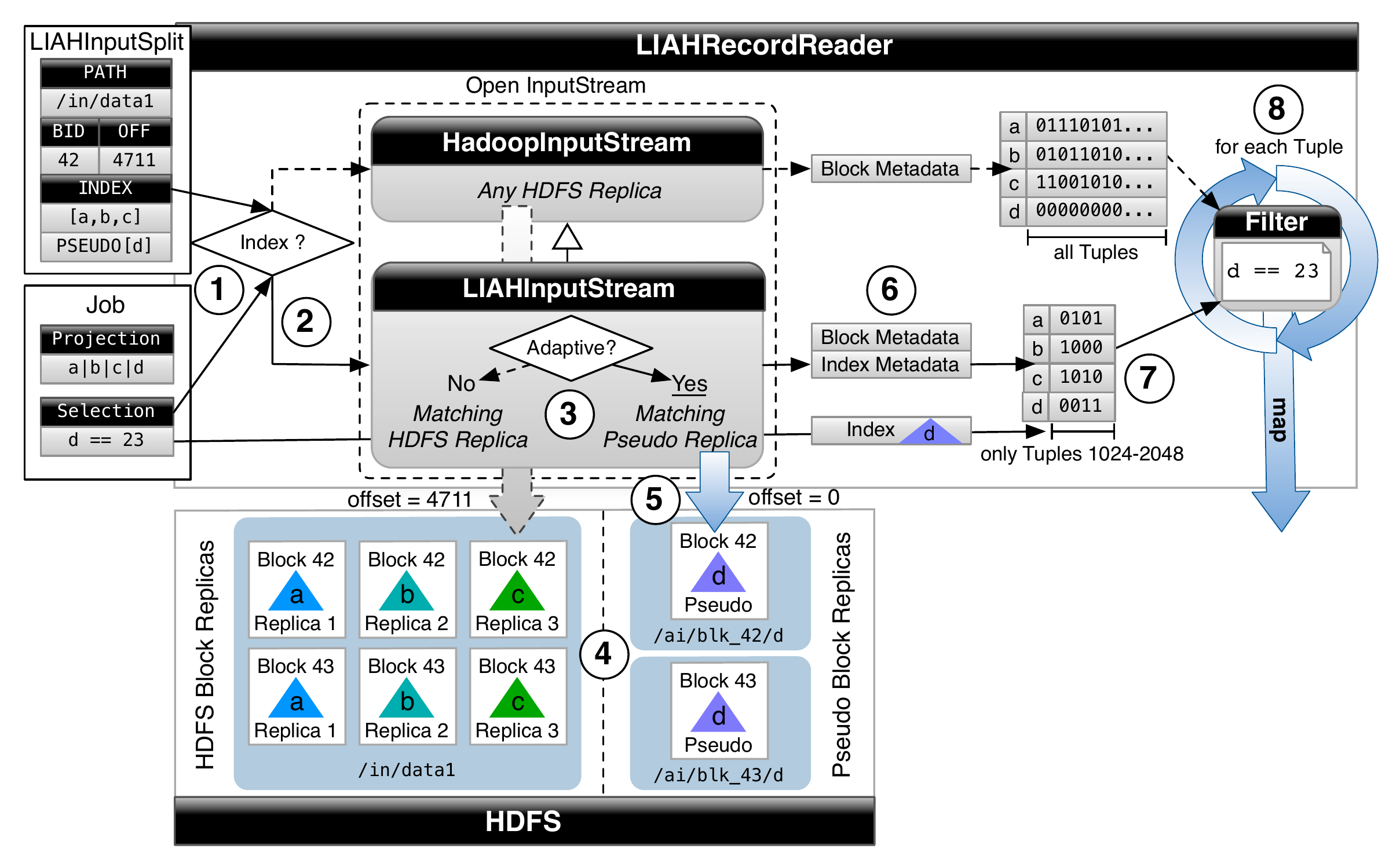}
\vspace{-0.3cm}
\caption{\hail\ RecordReader internals.}
\label{figure:maptask_internals}
\vspace{-0.3cm}
\end{figure*}

To solve this problem, the Index Writer creates a {\it pseudo data block replica}, which is a new HDFS file~\no{8}. Therefore, although the pseudo data block replica is a logical copy of $block_{42}$, the NameNode does not recognise it as a normal data block replica. Instead, the NameNode simply sees the pseudo data block replica as another index available for $block_{42}$. To avoid shipping data from one node to another, the Index Writer aims at storing the pseudo data block replica locally on DataNode~5. With this aim, the Index Writer stores the pseudo data block replica with replication factor one. The Index Writer follows a naming convention, which contains the identifier of the block and the main index attribute, to uniquely identify a pseudo data block replica. It is worth noting that a map task can compete with a speculative map task (speculative execution) in a different node to create a pseudo data block replica. To deal with this race condition, we follow the same pattern used by map tasks to store their intermediate output in original Hadoop. This means that each Index Writer first stores the pseudo data block replica in a temporary file and then tries to rename it after completion. Only the first Index Writer will succeed and the second one will remove its temporary pseudo data block replica. Additionally, the successful Index Writer informs the NameNode about the new index on attribute $d$~\no{9}. This allows \hail\ to take the newly created indexes into account for processing future jobs. 

\subsection{Pseudo Data Block Replicas}
\label{accessing_blocks}

Recall from Section~\ref{section:adaptive-indexer_indexwriter} that a pseudo data block replica is a logical copy of a data block (in a different sort order) that is stored by \hail\ as a new HDFS file rather than as a normal data block replica. This allows \hail\ to keep a different replication factor on a block basis rather than on a file basis. As pseudo data block replicas are stored in different HDFS files than normal data block replicas, an important question arises: {\it how to access pseudo data block replicas in an invisible way for users?}

\hail\ achieves this transparency via its RecordReader (the \hail\ RecordReader). Users continue annotating their map functions (with selection predicates and projections) as with HAIL. The \hail\ RecordReader takes care of automatically switching from normal to pseudo data block replicas. For this, the \hail\ RecordReader uses the {\it \hail\ InputStream}, a wrapper of the Hadoop FSInputStream. We discuss the details of both the \hail\ RecordReader and the \hail\ InputStream below.

Figure~\ref{figure:maptask_internals} illustrates the internal pipeline of the \hail\ RecordReader when processing a given \hail\ InputSplit. When a map task starts, the \hail\ RecordReader first reads the metadata of its \hail\ InputSplit in order to check if there exists a suitable index to process the input data block ($block_{42}$)~\no{1}. If a suitable index is available, the \hail\ RecordReader initialises the \hail\ InputStream with the selection predicate of $job_d$ as a parameter~\no{2}. Internally, the \hail\ InputStream checks if the index resides in a normal or pseudo data block replica~\no{3}. This allows the \hail\ InputStream to open an input stream to the right HDFS file. This is because normal and pseudo data block replicas are stored on different HDFS files. While all normal data block replicas belong to the same HDFS file, each pseudo data block replica belongs to a different HDFS file~\no{4}. As in our example the index on attribute $d$ for $block_{42}$ resides in a pseudo data block replica, the \hail\ InputStream opens an input stream to the HDFS file \texttt{/pseudo/blk\_42/d}~\no{5}. As a result, the \hail\ RecordReader does not care from which file it is reading since normal and pseudo data block replicas have the same format. Therefore, switching between a normal and a pseudo data block replica is not only invisible to users, but also to the \hail\ RecordReader. The \hail\ RecordReader just reads the block and index metadata using the \hail\ InputStream~\no{6}. After performing  an index lookup for the selection predicate of $job_d$, the \hail\ RecordReader loads only the qualifying tuples (e.g.~tuples 1024 -- 2048) from the projected attributes ($a$, $b$, $c$, and $d$)~\no{7}. Finally, the \hail\ RecordReader forms key-value pairs and passes only qualifying pairs to the map function~\no{8}. 

In case that no suitable index exists, the \hail\ RecordReader takes the Hadoop InputStream, which opens an input stream to any normal data block replica, and falls back to full scan.

\subsection{Lazy Indexing}
\label{incremental_indexing}
The blocking queues used by the Adaptive Indexer allow us to easily protect \hail\ against CPU overloading. However, writing pseudo data block replicas can also slow down the parallel read and write processes of MapReduce jobs. In fact, the negative impact of extra I/O operations can be high as MapReduce jobs are typically I/O-bound. As a result, \hail\ as a whole might become slower even if the Adaptive Indexer can computationally keep up with the job execution. So, the question that arises is: {\it how to write pseudo data block replicas efficiently?}

\hail\ solves this problem by making indexing incremental, i.e.~\hail\ spreads index creation over multiple MapReduce jobs. The goal is to balance index creation cost over multiple MapReduce jobs so that users perceive small (or no) overhead in their jobs. To do so, \hail\ uses an {\it offer rate}, which is a ratio that limits the maximum number of pseudo data block replicas (i.e.~number of data blocks to index) to create during a single MapReduce job. For example, using an offer rate of 10\%, \hail\ indexes in a single MapReduce job at maximum one data block out of ten processed data blocks (i.e.~\hail\ only indexes 10\% of the total data blocks). Notice that, consecutive jobs with selections on the same attribute benefit from pseudo data block replicas created during previous jobs. This strategy brings two major advantages. First, \hail\ can reduce the additional I/O introduced by indexing to any level that is acceptable for the user. Second, the indexing effort done by \hail\ is according to the current query workload. Another advantage of using an offer rate is that users can decide how fast they want to converge to a {\it complete index}, i.e.~all data blocks are indexed. For instance, using an offer rate of 10\%, \hail\ would require $10$ MapReduce jobs with a selection predicate on the same attribute to converge to a {\it complete index} (i.e.~all data blocks are indexed). Therefore, on the one hand, the investment in terms of time and space for MapReduce jobs with selection predicates on unfrequent attributes is minimised. However, on the other hand, MapReduce jobs with selection predicates on frequent attributes quickly converge to a completely indexed copy. We discuss more details about different offer rate strategies in Section~\ref{section:indexing_strategies}.


\subsection{\hail\ Scheduling}
\label{section:hail_scheduling}
An interesting result we found in~\cite{hail} is that the initialisation and finalisation costs of map tasks are so high that they basically dominate short running jobs. Thus, reducing the number of map tasks is crucial to improve the performance of MapReduce jobs.

To deal with this problem, we introduce \hail\ Scheduling, an extension of the HAIL Scheduling proposed in~\cite{hail}. \hail\ Scheduling works as follows. First, \hail\ Scheduling partitions all input data blocks into indexed data blocks and unindexed data blocks. Second, \hail\ Scheduling combines several indexed data blocks into one split as described in~\cite{hail}. This allows \hail\ to reduce the number of map tasks to schedule, thereby reducing the total overhead for initialising and finalising map tasks. Notice that, after this step, \hail\ can obtain the exact number of already existing indexes for each computing node. Third, like original Hadoop, \hail\ creates one map task per unindexed data block. For each map task, \hail\ considers $r$ different computing nodes as possible locations to schedule a map task, where $r$ is the replication factor of the input dataset. However, in contrast to original Hadoop, \hail\ tries to schedule a map task to the computing node with the smallest number of existing indexes. As a result, \hail\ can: (i)~better parallelise index access for future MapReduce jobs and (ii)~increase the chances to keep both normal and pseudo data block replicas in the same node. The last point prevents \hail\ to shuffle data through the network when writing pseudo data block replicas.

%
%

%
%
%
%
%
%
%
%
%

\section{Adaptive Indexing Strategies}
\label{section:indexing_strategies}
We now present three strategies that allow \hail\ to improve the performance of MapReduce jobs. We first present {\it eager indexing}, a technique that allows \hail\ to adapt its incremental indexing mechanism to the number of already created pseudo data block replicas. We then discuss how \hail\ can prioritise data blocks for indexing based on their selectivity. Finally, we introduce {\it invisible projection}, a new technique to deal with partial projections of users' MapReduce jobs.

\subsection{Eager Adaptive Indexing}
\label{section:eager}
Recall that \hail\ uses an {\it offer rate} to throttle down adaptive indexing efforts to an acceptable (or even invisible) degree for users (see Section~\ref{incremental_indexing}). However, let us make two important observations that could make a constant offer rate not desirable for certain users:

\vspace{-0.2cm}
\begin{packed_enum}
\item Using a constant offer rate, the job runtime of consecutive MapReduce jobs having a filter condition on the same attribute is not constant. Instead, they have an almost linearly decreasing runtime up to the point where all blocks are indexed. This is because the first MapReduce job is the only to perform a full scan over all the data blocks of a given dataset. Consecutive jobs, even when indexing and storing the same amount of blocks, are likely to run faster as they benefit from all indexing work of their predecessors.

\item \hail\ actually delays indexing by using an offer rate. The tradeoff here is that using a lower offer rate leads to a lower indexing overhead, but it requires more MapReduce jobs to index all the data blocks in a given dataset. However, some users want to limit the experienced indexing overhead and still desire to benefit from complete indexing as soon as possible.
\end{packed_enum}



\begin{table}[t!]
\vspace{-0.2cm}
\caption{\label{table:notations}Cost model parameters.}
\begin{center}
\begin{tabular}{l p{6cm}}
\hline
{\bf Notation} & {\bf Description} \\
\hline
$n_{slots}$ & The number of map tasks that can run in parallel in a given Hadoop cluster \\
$n_{blocks}$ & The number of data blocks of a given dataset \\
$n_{idxBlocks}$ & The number of blocks with a relevant index \\
$n_{fsw}$ & The number of map waves performing a full scan\\
$t_{fsw}$ & The {\it average} runtime of a map wave performing a full scan (without adaptive indexing overhead) \\
$t_{idxOverhead}$ & The average time overhead of adaptive indexing in a map wave \\
$T_{idxOverhead}$ & The {\it total} time overhead of adaptive indexing \\
$T_{is}$ & The total runtime of the map waves performing an index scan \\
$T_{job}$ & The total runtime of a given job \\
$T_{target}$ & The targeted total job runtime \\
$\rho$ & The ratio of data blocks (w.r.t.~ $n_{blocks}$) offered to the Adaptive Indexer \\
\hline
\end{tabular}
\end{center}
\vspace{-0.5cm}
\end{table}

\vspace{-0.2cm}
Therefore, we propose an \textit{eager adaptive indexing} strategy to deal with this problem. The basic idea of eager adaptive indexing is to dynamically adapt the offer rate for MapReduce jobs according to the indexing work achieved by previous jobs. In other words, eager adaptive indexing tries to exploit the saved runtime and reinvest it as much as possible into further indexing. To do so, \hail\ first needs to estimate the runtime gain (in a given MapReduce job) from performing an index scan on the already created pseudo data block replicas. For this, \hail\ uses a cost model to estimate the total runtime, $T_{job}$, of a given MapReduce job (Equation~\ref{equation:jobRT}). Table~\ref{table:notations} lists the parameters we use in the cost model.
\begin{equation}
\label{equation:jobRT}
T_{job} = T_{is} + t_{fsw}\cdot n_{fsw} + T_{idxOverhead}
\end{equation}
We define the number of map waves performing a full scan, $n_{fsw}$, as $\lceil\frac{n_{blocks}-n_{idxBlocks}}{n_{slots}}\rceil$.
Intuitively, the total runtime $T_{job}$ of a job consists of three parts. First, the time required by \hail\ to process the existing pseudo data block replicas, i.e.~all data blocks having a relevant index, $T_{is}$. Second, the time required by \hail\ to process the data blocks without a relevant index, $t_{fsw}\cdot n_{fsw}$. Third, the time overhead caused by adaptive indexing, $T_{idxOverhead}$.\footnote{It is worth noting that $T_{idxOverhead}$ denotes only the additional runtime that a MapReduce job has due to adaptive indexing.} The adaptive indexing overhead depends on the number of data blocks that are offered to the Adaptive Indexer and the average time overhead observed for indexing a block. Formally, we define $T_{idxOverhead}$ as follows:
\begin{equation}
\label{equation:idxOverhead}
T_{idxOverhead} = t_{idxOverhead} \cdot \min\left( \rho \cdot \left\lceil\textstyle\frac{n_{blocks}}{n_{slots}}\right\rceil ,n_{fsw} \right)
\end{equation}

We can use this model to automatically calculate the offer rate $\rho$ in order to keep the adaptive indexing overhead acceptable for users. Formally, from Equations~\ref{equation:jobRT} and~\ref{equation:idxOverhead}, we deduct $\rho$ as follows: 
\begin{equation*}
\label{equation:rho}
\rho =\frac{T_{target} -  T_{is} - t_{fsw}\cdot n_{fsw} }{ t_{idxOverhead} \cdot \lceil\textstyle\frac{n_{blocks}}{n_{slots}}\rceil}
\end{equation*}

Therefore, given a target job runtime $T_{target}$, \hail\ can automatically set $\rho$ in order to fully use this time budget for creating indexes and use the gained runtime in the next jobs either to speed up the jobs or to create even more indexes. Usually, we choose $T_{target}$ to be equal to the runtime of the very first job so that users can observe a stable runtime till almost everything is indexed. However, users can set $T_{target}$ to any time budget in order to adapt the indexing effort to their needs. Notice that, since accessing pseudo data block replicas is independent of $\rho$, \hail\ first processes pseudo data block replicas and measures $T_{is}$, before deciding what offer rate to use for the unindexed blocks. The average runtimes $t_{fsw}$ (from Equation~\ref{equation:jobRT}) and $t_{idxOverhead}$ (from Equation~\ref{equation:idxOverhead}) can be measured in a calibration job or given by users.

On the one hand, \hail\ can now adapt the offer rates to the performance gains obtained from performing index scans over the already indexed data blocks. On the other hand, by gradually increasing the offer rate, eager adaptive indexing prioritises complete index convergence over early runtime improvements for users. Thus, users no longer experience an incremental and linear speed up in job performance until the index is eventually complete, but instead they experience a sharp improvement when \hail\ approaches to a complete index. In summary, besides limiting the overhead of adaptive indexing, the offer rate can also be considered as a tuning knob to trade early runtime improvements with faster indexing.

\subsection{Selectivity-based Indexing}
Earlier, we saw that \hail\ uses an offer rate to limit the number of data blocks to index in a single MapReduce job. For this, \hail\ uses a round robin policy to select the data blocks to pass to the Adaptive Indexer. This sounds reasonable under the assumption that data is uniformly distributed. However, datasets are typically skewed in practice and hence some data blocks might contains more qualifying tuples than others under a given query workload. Consequently, indexing highly selective data blocks before other data blocks promises higher performance benefits.

Therefore, \hail\ can also use a selectivity-based data block selection approach for deciding which data blocks to use. The overall idea is to use available computing resources in order to maximise the expected performance improvement for future MapReduce jobs running on partially indexed datasets. The big advantage of this approach is that users can perceive higher improvements in performance for their MapReduce jobs from the very first runs. Additionally, as a side-effect of using this approach, \hail\ can adapt faster to the selection predicates of MapReduce jobs. However, {\it how can \hail\ efficiently obtain the selectivities of data blocks?}

For this, \hail\ exploits the natural process of map tasks to propose data blocks to the Adaptive Indexer. Recall that a map task passes a data block to the Adaptive Indexer once the map task finished processing the block. Thus, \hail\ can obtain the accurate selectivity of a data block by piggybacking on the map phase: when the data block is filtered according to the provided selection predicate. This allows \hail\ to have perfect knowledge about selectivities for free. Given the selectivity of a data block, \hail\ can decide if it is worth to index the data block or not. In our current \hail\ prototype, a map task proposes a data block to the Adaptive Indexer if the percentage of qualifying tuples in the data block is equal or higher than 80\%. However, users can adapt this threshold to their applications. Notice that with the statistics on data block selectivities, \hail\ can also decide which indexes to drop in case of storage limitations. However, a discussion on an index eviction strategy is out of the scope of this paper.

\subsection{Invisible Projection}
\label{section:invisible_projection}
In Section~\ref{section:adaptivehail_pipeline}, we discussed the general flow of \hail\ for creating clustered indexes in an adaptive and incremental manner. For simplicity, in such discussion, we implicitly assume that consecutive MapReduce jobs (with a selection predicate on the same attribute) read all (or the same) attributes from disk. Indeed, this is the simplest case for indexing data blocks, because the entire data blocks (or all required attributes) are available in main memory. As a result, the index attribute can be sorted and all other attributes can be reordered at the same time for alignment reasons.

However, this becomes challenging when MapReduce jobs have different projections. For example, consider a dataset having four attributes \textit{a}, \textit{b}, \textit{c}, \textit{d}. Assume that \hail\ is using an offer rate of 50\% and that there is no index on any of these attributes yet. Now, consider a first MapReduce job (say $job_1$) having a selection predicate on $d$ and projecting attribute $b$. Since \hail\ does not have an index on attribute $d$, \hail\ creates a clustered index on attribute $d$ and aligns attribute $b$ with respect to attribute $d$. As a result of running $job_1$, \hail\ ends up by having a clustered index on $d$ (and $b$ aligned) for 50\% of the data blocks (one index for each block) in the dataset. Now, consider a second MapReduce job (say $job_2$) having a selection predicate on $d$ and projecting attribute $c$. It is here that \hail\ faces a problem: \hail\ cannot fully benefit from the already 50\% indexed data blocks in the dataset. This is because attribute $c$ is still not aligned with respect to attribute $d$. Hence, performing an index scan for these data blocks would lead \hail\ to perform many random I/O operations for tuple reconstruction. The same applies for other MapReduce jobs with different projections. For instance, a MapReduce job (say $job_3$), having a selection predicate on attribute $d$ and projecting attribute $a$, cannot benefit from the indexes created neither by $job_1$ nor by $job_2$. Therefore, the question is: {\it how to create clustered indexes when MapReduce jobs have filter conditions on the same attribute but have different projections?}

To deal with this problem, we introduce the {\it invisible projection} technique. The idea is to additionally read all missing attributes before passing a data block to the Adaptive Indexer. Notice that, \hail\ applies the invisible projection technique only for the data blocks that map tasks propose to the Adaptive Indexer. The beauty of this technique is that it is transparent for users as map tasks still process the attributes required by MapReduce jobs. For example, consider again $job_1$ from the above example. In this case, \hail\ (the HAIL RecordReader) provides only attributes $d$ and $b$ to the map function of map tasks, but \hail\ provides all attributes ($a$, $b$, $c$, and $d$) to the Adaptive Indexer. This way \hail\ ensures that data blocks are always completely available in main memory for the Adaptive Indexer.

The reader might think that the invisible projection approach is not suitable for data-intensive applications as it requires still extra I/O operations for loading unprojected attributes. However, this is far from the truth, because the extra I/O operations are only for the data blocks offered to the Adaptive Indexer. Furthermore, the impact of invisible projection also depends on the proportion of projected attributes. Therefore, if the proportion of projected attributes is low, \hail\ can always decrease the offer rate in order to keep an acceptable adaptive indexing overhead. Alternatively, \hail\ could also switch to a {\it lazy projection} approach to only read the attributes required by MapReduce jobs. Even though the current \hail\ prototype supports only the invisible projection technique, we discuss the lazy projection technique in Appendix~\ref{appendix:lazy_reordering}.

\section{Experiments}
\label{experiments}
We evaluate the efficiency of \hail\ to adapt to query workloads and compare it with Hadoop and HAIL. We measure the performance of \hail\ with three main objectives in mind: (i)~to measure the adaptive indexing overhead that \hail\ generates over the runtime of MapReduce jobs; (ii)~to evaluate both how fast \hail\ can adapt to workloads and how well MapReduce jobs can benefit from \hail; (iii)~to study how well each of the adaptive indexing strategies of \hail\ allow MapReduce jobs to improve their runtime.

\subsection{Setup}
\label{section:exp_setup}

\vspace{0.1cm}
\noindent{\bf Cluster.}
We use two different clusters in our experiments. Our first cluster ({\it Cluster-A}), is a 10-node cluster where each node has: one 2.66GHz Quad Core Xeon processor; 4x4GB of main memory; 1x750GB SATA hard disk; three one Gigabit network cards. Our second cluster ({\it Cluster-B}), is a 4-node cluster where each node has: one 3.46 GHz Hexa Core Xeon X5690 processors; 20GB of main memory; one 278GB SATA hard disk (for the OS) and one 837GB SATA hard disk (for HDFS); two one Gigabit network cards. We use Cluster-B to measure the influence of more efficient processors on the behavior of \hail. For both Cluster-A and Cluster-B, we use a 64-bit openSUSE 12.1 OS and the ext3 filesystem.

\vspace{0.1cm}
\noindent{\bf Datasets.}
We use the web log dataset (\texttt{UserVisits}) from the HAIL paper~\cite{hail}. This dataset has nine attributes, which are mostly strings, and has a total size of $40GB \times${\it numberOfNodes}, i.e.~400GB for Cluster-A and 160GB for Cluster-B. Additionally, we use a \texttt{Synthetic} dataset containing only numeric attributes as scientific datasets. The \texttt{Synthetic} dataset has six attributes and a total size of $50GB \times${\it  numberOfNodes}, i.e.~500GB for Cluster-A and 200GB for Cluster-B. We generate the values for the first attribute in the range $[1..10]$ and with an exponential repetition for each value, i.e.~$10^{i-1}$ where $i\in[1..10]$. We generate the other five attributes at random. Then, we shuffle all tuples across the entire dataset in order to have the same distribution across data blocks.

\vspace{0.1cm}
\noindent{\bf MapReduce Jobs.}
For the \texttt{UserVisits} dataset, we consider eleven jobs (\texttt{JobUV1} -- \texttt{JobUV11}) with a selection predicate on attribute \texttt{searchWord} and with a full projection (i.e.~projecting all $9$ attributes). The first four jobs \texttt{JobUV1} -- \texttt{JobUV4} have a selectivity of $0.4\%$ ($1.24$ million output records) and the remaining seven jobs (\texttt{JobUV5} -- \texttt{JobUV11}) have a selectivity of $0.2\%$ ($0.62$ million output records). For the \texttt{Synthetic} dataset, we consider other eleven jobs (\texttt{JobSyn1} -- \texttt{JobSyn11}) with a full projection, but with a selection predicate on the first attribute. These jobs have a selectively of $0.2\%$ ($2.2$ million output records). All MapReduce jobs for both datasets select disjoint ranges to avoid caching effects. For all experiments, we report the average of three trials.

\vspace{0.1cm}
\noindent{\bf Systems.}
We use Hadoop v0.20.203 and HAIL as baseline systems to evaluate the benefits of \hail . For Hadoop, we use the default configuration settings, but increase the data block size to 256MB to decrease the scheduling overhead for Hadoop. In our experiments using the \texttt{UserVisits} dataset, HAIL creates one index for attribute \texttt{sourceIP}, one for attribute \texttt{visitDate}, and one for attribute \texttt{adRevenue}, just like in~\cite{hail}. For our experiments using the \texttt{Synthetic} dataset, we simply assume that HAIL does not create an index on the first attribute. For \hail\, we consider four different variants according to the offer rate ($\rho$) we use: \hail\ ($\rho=0.1$), \hail\ ($\rho=0.25$), \hail\ ($\rho=0.5$), and \hail\ ($\rho=1$). 

\newpage

\subsection{Performance for the First Job}
\label{section:exp_overhead}
Since \hail\ piggybacks adaptive indexing on MapReduce jobs, the very first question that the reader might ask is: {\it what is the additional runtime incurred by \hail\ on MapReduce jobs?} We answer this question in this section. For this, we run job \texttt{JobUV1} for the \texttt{UserVisits} dataset and job \texttt{JobSyn1} for the \texttt{Synthetic} datasets. For these experiments, we assume that there is no block with a relevant index for jobs \texttt{JobUV1} and \texttt{JobSyn1}.

\begin{figure}[!t]
\centering\includegraphics[width=\linewidth]{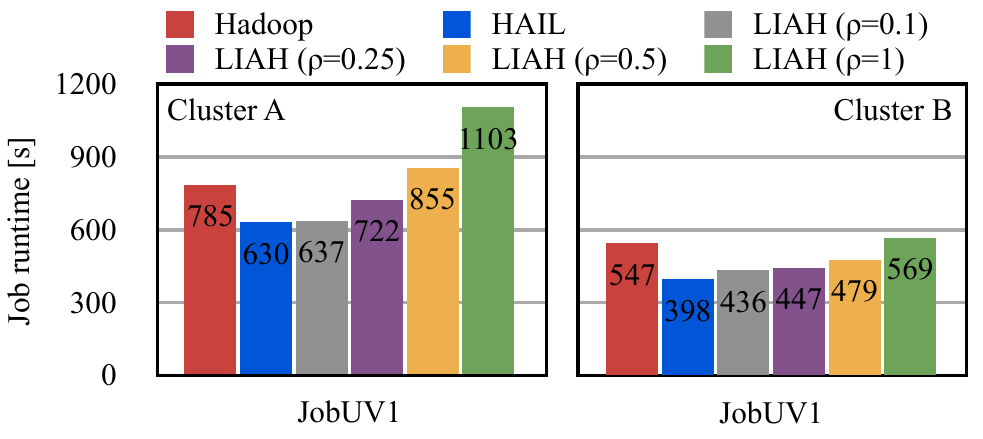}
\vspace{-0.8cm}
\caption{\hail\ Performance when running the first  MapReduce job over \texttt{UserVisits}.}
\label{figure:offerOverheadPavlo}
\vspace{-0.3cm}
\end{figure}

Figure~\ref{figure:offerOverheadPavlo} shows the job runtime for the four variants of \hail\ for the \texttt{UserVisits} dataset. In Cluster-A, we observe that \hail\ has almost no overhead (only 1\%) over HAIL when using an offer rate of 10\% (i.e.~$\rho=0.1$). Interestingly, we observe that \hail\ is still faster than Hadoop with $\rho=0.1$ and $\rho=0.25$. Indeed, the overhead incurred by \hail\ increases along with the offer rate used by \hail . However, we observe that \hail\ increases the execution time of \texttt{JobUV1} by less than factor of two w.r.t. both Hadoop and HAIL, even though all data blocks are indexed in a single MapReduce job. We especially observe that the overhead incurred by \hail\ scales linearly with the ratio of indexed data blocks (i.e.~with $\rho$), except when scaling from $\rho=0.1$ to $\rho=0.25$. This is because \hail\ starts to be CPU bound only when offering more than 20\% of the data blocks (i.e.~from $\rho=0.25$). This changes when running \texttt{JobUV1} in Cluster-B. In these results, we clearly observe that the overhead incurred by \hail\ scales linearly with $\rho$. We especially observe that \hail\ benefits from using newer CPUs and have better performance than Hadoop for most offer rates. \hail\ has only $4\%$ overhead over Hadoop when having $\rho=1$. Additionally, we can see that \hail\ has low overheads w.r.t. HAIL: from $10\%$ (with $\rho=0.1$) to $43\%$ (with $\rho=1$). 

\begin{figure}[!t]
\centering\includegraphics[width=\linewidth]{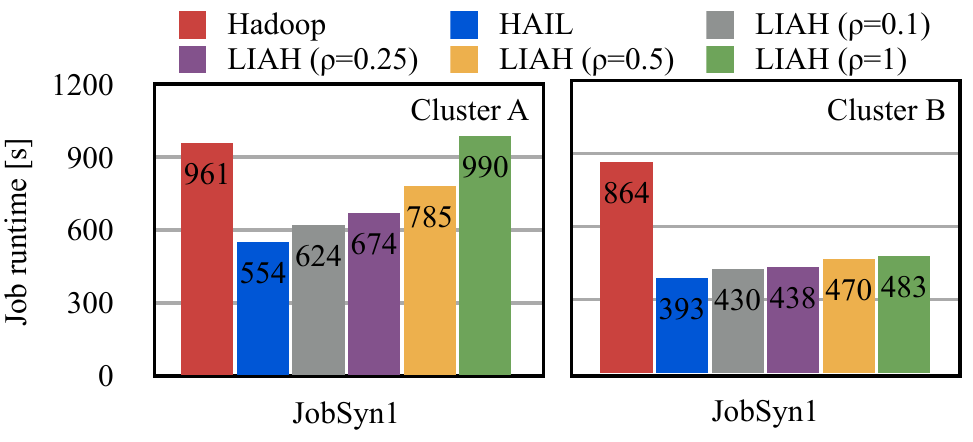}
\vspace{-0.8cm}
\caption{\hail\ Performance when running the first  MapReduce job over \texttt{Synthetic}.}
\label{figure:offerOverheadSyn}
\vspace{-0.4cm}
\end{figure}

Figure~\ref{figure:offerOverheadSyn} shows the job runtimes for \texttt{Synthetic}. Overall, we observe that the overhead incurred by \hail\ continues to scale linearly with the offer rate. In particular, we observe that \hail\ has no overhead over Hadoop in both clusters, except for \hail\ ($\rho=1$) in Cluster-A (where \hail\ incurs a negligible overhead of $\sim\!\!3\%$). It is worth noting that when using newer CPUs (Cluster-B) \hail\ has very low overheads over HAIL as well: from $9\%$ to only $23\%$.

From these results, we can conclude that \hail\ can efficiently create indexes at job runtime while limiting the overhead of writing pseudo data block replicas. In particular, we observe the efficiency of the lazy indexing mechanism of \hail\ to adapt to users' requirements via different offer rates.

\subsection{Performance for a Sequence of Jobs}
\label{section:exp_sequence}
We saw in the previous section that \hail\ can linearly scale the adaptive indexing overhead with the help of the offer rate. But, {\it which are the implications for a sequence of MapReduce jobs?} To answer this question, we run the sequence of eleven MapReduce jobs for each dataset: \texttt{JobUV1} -- \texttt{JobUV11} for \texttt{UserVisits} and \texttt{JobSyn1}~-- \texttt{JobSyn11} for \texttt{Synthetic}.
 
\begin{figure}[!t]
\centering\includegraphics[width=\linewidth]{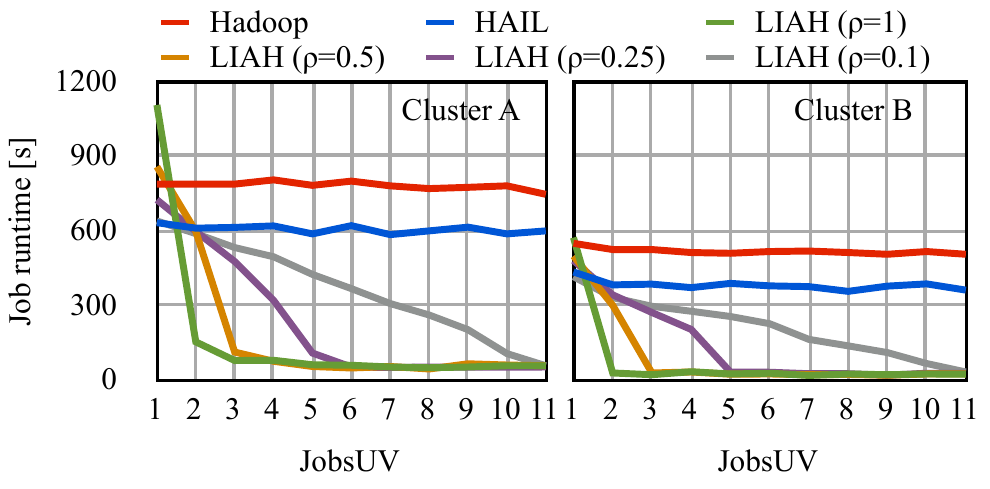}
\vspace{-0.8cm}
\caption{\hail\ performance when running a sequence of MapReduce jobs over \texttt{UserVisits}.}
\label{figure:offerPavlo}
\vspace{-0.3cm}
\end{figure}

\begin{figure}[!t]
\centering\includegraphics[width=\linewidth]{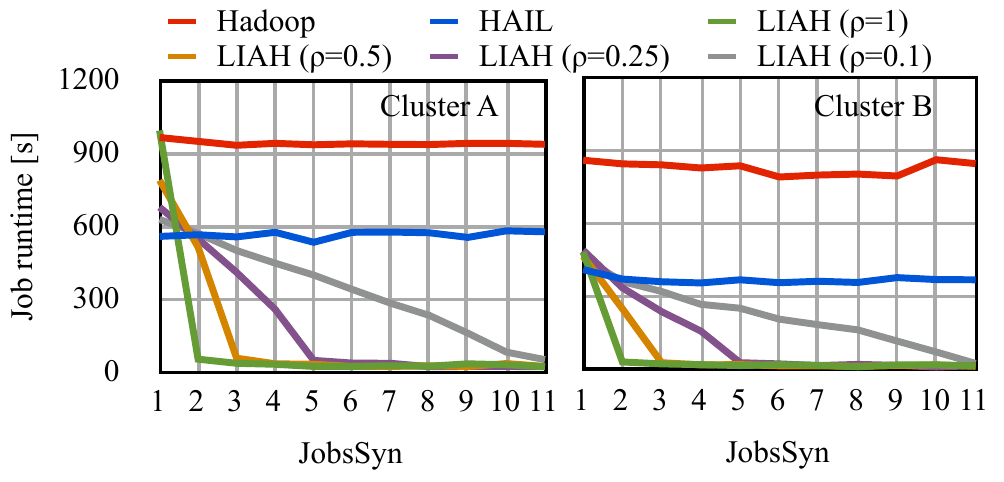}
\vspace{-0.8cm}
\caption{\hail\ performance when running a sequence of MapReduce jobs over \texttt{Synthetic}.}
\label{figure:offerSyn}
\vspace{-0.3cm}
\end{figure}

Figures~\ref{figure:offerPavlo} and~\ref{figure:offerSyn} show the job runtimes for the \texttt{UserVisit} and \texttt{Synthetic} datasets, respectively. Overall, we clearly see in both computing clusters that \hail\ improves the performance of MapReduce jobs linearly with the number of indexed data blocks. In particular, we observe that the higher the offer rate, the faster \hail\ converges to a complete index. However, the higher the offer rate, the higher the adaptive indexing overhead for the initial job (\texttt{JobUV1} and \texttt{JobSyn1}). Thus, users are faced with a natural tradeoff between indexing overhead and the required number of jobs to index all blocks. But, it is worth noting that users can use low offer rates (e.g.~$\rho=0.1$) and still quickly converge to a complete index (e.g.~after $10$  job executions for $\rho=0.1$). In particular, we observe that after executing only a few jobs \hail\ already outperforms Hadoop and HAIL significantly. For example, let us consider the sequence of jobs on \texttt{Synthetic} using $\rho=0.25$ on Cluster-B. Remember that for this offer rate the overhead for the first job compared to HAIL is relatively small (11\%) while \hail\ is still able to outperform Hadoop. With the second job \hail\ is slightly faster than HAIL and when running the fourth job improves over  HAIL by more than a factor of two and over Hadoop by more than a factor of five\footnote{Although \hail\ is still indexing further blocks.}. 
 As soon as \hail\ converges to a complete index, \hail\ significantly outperforms HAIL by up to a factor of $23$ and Hadoop by up to a factor of $52$. For the \texttt{UserVisits} dataset, \hail\ outperforms  HAIL by up to a factor of $24$ and Hadoop by up to a factor of $32$. 
 Notice that, \hail\ and HAIL increase the performance gap with Hadoop for \texttt{Synthetic}, because they significantly reduce the size of this dataset when converting it to binary representation.

In summary, the results show that \hail\ can efficiently adapt to query workloads with a very low overhead only for the very first job: the following jobs always benefit from the indexes created in previous jobs. Interestingly, an important result is that \hail\ can converge to a complete index after running only a few jobs.

\newpage

\subsection{Eager Adaptive Indexing for a Sequence of Jobs}
\label{section:exp_eager}
We saw in the previous section that \hail\ improves the performance of MapReduce jobs linearly with the number of indexed data blocks. Now, the question that might arise in the reader's mind is: {\it can \hail\ efficiently exploit the saved runtimes for further adaptive indexing?} To answer this question, we enable the eager adaptive indexing strategy in \hail\ and run again all \texttt{UserVisits} jobs using an initial offer rate of $10\%$. In these experiments, we use Cluster-A and consider \hail\ (without eager adaptive indexing enabled) with offer rates of $10\%$ and $100\%$ as baselines.

\begin{figure}[!t]
\centering\includegraphics[width=\linewidth]{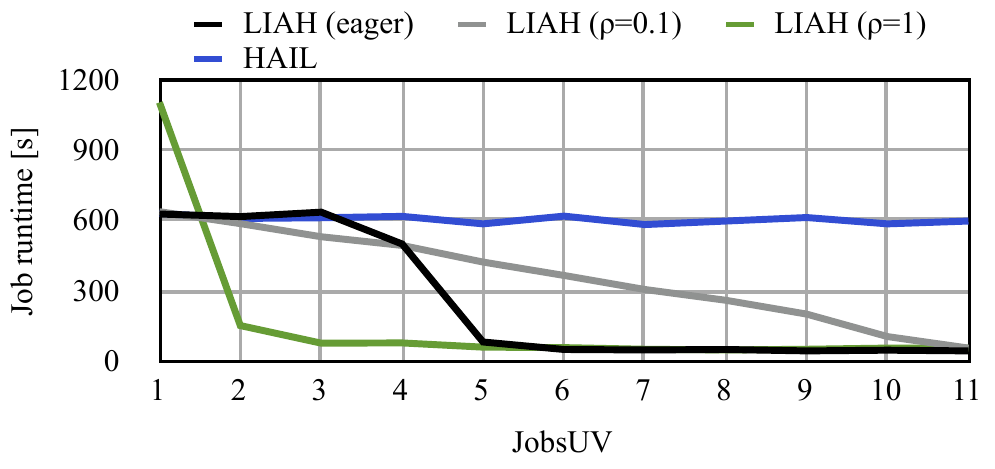}
\vspace{-0.6cm}
\caption{Eager adaptive indexing vs. $\rho = 0.1$ and $\rho=1$}
\label{figure:eagerIndexing}
\vspace{-0.3cm}
\end{figure}

Figure~\ref{figure:eagerIndexing} show the result of this experiment. As expected, we observe that \hail\ (eager) has the same performance as \hail\ ($\rho=0.1$) for \texttt{JobUV1}. However, in contrast to \hail\ ($\rho=0.1$), \hail\ (eager) keeps its performance constant for \texttt{JobUV2}. 
This is because \hail\ (eager) automatically increases $\rho$ from $0.1$ to $0.17$ in order to exploit saved runtimes. For \texttt{JobUV3}, \hail\ (eager) still keeps its performance constant by increasing $\rho$ from $0.17$ to $0.33$. Now, even though \hail\ (eager) increases $\rho$ from $0.33$ to $1$ for \texttt{JobUV4}, \hail\ (eager) now improves the job runtime as only $40\%$ of the data blocks remain unindexed. As a result of adapting its offer rate, \hail\ (eager) converges to a complete index only after $4$ jobs while incurring almost no overhead over HAIL. From \texttt{JobUV5}, \hail\ (eager) ensures the same performance as \hail\ ($\rho=1$) since all data blocks are indexed, while \hail\ ($\rho=0.1$) takes $6$ more jobs to converge to a complete index.

These results show that \hail\ can converge even faster to a complete index, while still keeping a negligible indexing overhead for users' MapReduce jobs. Overall, these results demonstrate the high efficiency of \hail\ (eager) to adapt its offer rate according to the number of already indexed data blocks.

\subsection{Invisible Projection for the First Job}
\label{section:exp_invisible_projection}
So far, we have considered MapReduce jobs that project all attributes. However, this is not always the case in practice. Therefore, it is also important to answer the following question: {\it how well does \hail\ deal with MapReduce jobs that project only a subset of attributes from their input datasets?} We focus on answering this question in this section. To do so, we now enable the invisible projection technique from Section~\ref{section:invisible_projection} for \hail.

In these experiments, we consider two variants of \hail : one without invisible projection (\hail{\it woInvPrj}) and another with invisible projection (\hail{\it wInvPrj}).~We consider a constant offer rate of $25\%$ for these two variants of \hail . For both \texttt{UserVisits} and \texttt{Synthetic}, we run a MapReduce job with a selection predicate on the first attribute and vary the number of projected attributes. Overall, the main goal of these experiments is to measure the overhead of \hail{\it wInvPrj} (i.e.~reading all attributes) over \hail{\it woInvPrj} (i.e.~reading only the required attributes). To better evaluate the invisible projection technique, we assume that no index exists in \texttt{UserVisits} and \texttt{Synthetic}. We run these  experiments on Cluster-A.

\begin{figure}[!t]
\centering\includegraphics[width=\linewidth]{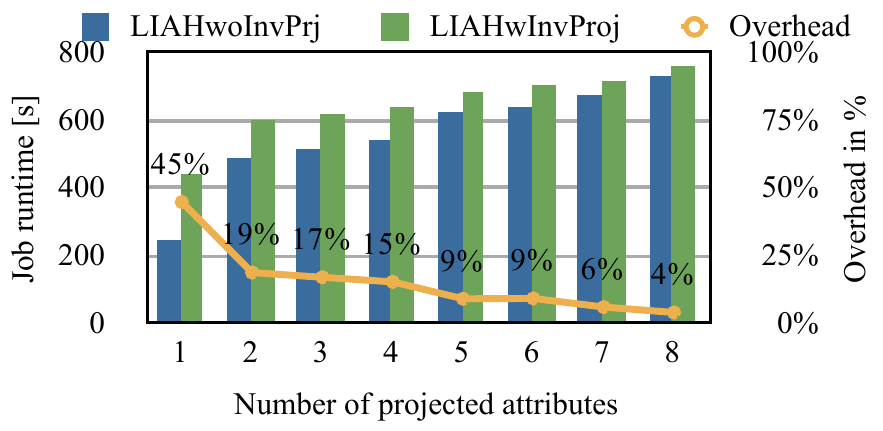}
\vspace{-0.7cm}
\caption{Invisible projection overhead for \texttt{UserVisits}.}
\label{figure:invisiblePavlo}
\vspace{-0.2cm}
\end{figure}

Figure~\ref{figure:invisiblePavlo} shows the results for \texttt{UserVisits}. We observe that, when \texttt{JobUV1} projects only the first attribute, \hail{\it wInvPrj} incurs an overhead of almost $45\%$. Indeed, this is partially because \hail{\it wInvPrj} has to read eight attributes more than \hail{\it woInvPrj}. However, most of this overhead is for reading the second attribute (\texttt{destURL}), which is the largest attribute in \texttt{UserVisits}. As soon as \hail{\it woInvPrj} also reads the second attribute, \hail{\it wInvPrj} incurs an overhead of only $\sim\!\!19\%$ (i.e.~$2x$ less overhead), even if \hail{\it wInvPrj} reads seven more attributes. Then, \hail{\it wInvPrj} lowers its overhead by $\sim\!\!2\%$ along with the number of projected attributes by \hail{\it woInvPrj}. But, as soon as \hail{\it woInvPrj} reads the second largest attribute (i.e.~the 5th attribute) in \texttt{UserVisits}, \hail{\it wInvPrj} again decreases its overhead by roughly a factor of $2$, i.e.~it now incurs an overhead of only $9\%$. From this point, the overhead caused by \hail{\it wInvPrj} starts to be negligible.

We saw in the results for \texttt{UserVisits} that \hail{\it wInvPrj} incurs a low overhead overall, especially as soon as \hail{\it woInvPrj} reads the most expensive attributes. Thus, the question that arises is: {\it how good is the invisible projection technique for scientific-like datasets, where most attributes are of the same size?} This is why we run again the invisible projection over the \texttt{Synthetic} dataset.

\begin{figure}[!t]
\centering\includegraphics[width=\linewidth]{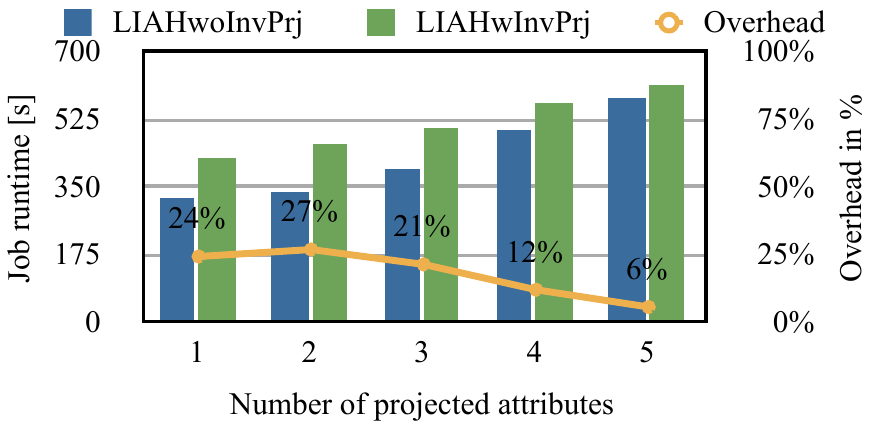}
\vspace{-0.7cm}
\caption{Invisible projection overhead for \texttt{Synthetic} with very high job selectivity.}
\label{figure:invisibleSyn_high}
\vspace{-0.3cm}
\end{figure}

Figure~\ref{figure:invisibleSyn_high} shows the results for \hail{\it wInvPrj} on the \texttt{Syntnhetic} dataset when having a very selective job (the job outputs only one tuple). Notice that, we consider a very selective job for this experiment in order to clearly see the impact of writing adaptively created indexes to disk. We observe that \hail{\it wInvPrj} has an acceptable overhead of $18\%$ on average and a low overhead when projecting more than the half of the total number of attributes. What it is interesting to highlight in these results is that the impact of the \hail{\it wInvPrj} over \hail{\it woInvPrj} is noticeable, because the MapReduce jobs outputs only a single tuple. To support this claim, we additionally ran a series of experiments with a very lowly selective job that outputs 80\% of the incoming tuples. In those experiments, we observed that \hail{\it wInvPrj} has a negligible $6\%$ overhead on average over \hail{\it woInvPrj}.

Additionally, we evaluated \hail{\it wInvPrj} on Cluster-B using an offer rate of 25\% and the \texttt{Synthetic} dataset, but we do not report the results here because of space constraints. Overall, we observed in this additional experiment that \hail{\it wInvPrj} incurs a negligible overhead of 1\% on average over \hail{\it woInvPrj}. These results also showed that \hail\ significantly benefits from using newer CPUs.

In summary, the results we presented in this section demonstrate the high efficiency of the invisible projection technique to deal with partial projections.




\section{Conclusion}
Several research works have improved the performance of MapReduce jobs significantly by integrating indexing into the MapReduce framework~\cite{traverse,fulltextindex,hadooppp,mapredPerf,hail}. However, none of these indexing techniques can adapt to changes in users' workload as they create indexes upfront. Therefore, these indexing techniques are not suitable for applications where workloads are hard to predict, such as in scientific applications and social networks.

In this paper, we proposed \hail\ (for {\it Lazy Indexing and Adaptivity in Hadoop}), a parallel, adaptive approach for indexing at minimal costs in MapReduce systems. \hail\ creates clustered indexes on data blocks as byproduct of MapReduce job execution. As a consequence, \hail\ can adapt to changes in user's workloads. The beauty of \hail\ is that it efficiently piggybacks index creation on the existing Hadoop MapReduce pipeline. Hence, \hail\ not only has no additional read I/O-costs, but it is also completely invisible for both the MapReduce system and users. A salient feature of \hail\ is that, besides distributing indexing effort across multiple computing nodes, it also parallelises indexing with map tasks computation and disk I/O. Furthermore, \hail\ can adjust the maximum number of data blocks to index in parallel with a single MapReduce job. In particular, we proposed {\it eager adaptive indexing}, a technique that allows \hail\ to reinvest the runtime benefits from indexes created by previous MapReduce jobs for further indexing. Thereby, \hail\ can trade the number of jobs to complete indexing a dataset with early job runtime improvements. This allows \hail\ to scale indexing effort according to hardware capabilities and users' needs. As a result, in contrast to existing adaptive indexing works, \hail\ incurs very low (or invisible) indexing overheads even for the first query that triggers the creation of a new index. Still, \hail\ quickly converges to a complete index, i.e.~all HDFS data blocks are indexed. Additionally, we introduced the {\it invisible projection} and {\it lazy projection} techniques, which allow \hail\ to efficiently create clustered indexes even if incoming jobs project only a subset of attributes from their input datasets.

We experimentally evaluated \hail\ and compared it with Hadoop and HAIL, using two different datasets (\texttt{UserVisits} and \texttt{Synthetic}) and computing clusters. The results demonstrated the high superiority of \hail: \hail\ runs MapReduce jobs up to $52$ times faster than Hadoop and up to $24$ times faster than HAIL. In particular, the results showed that \hail\ significantly outperforms Hadoop in almost all scenarios, except for the very first query when using an offer rate of 100\%. With respect to HAIL, \hail\ has a very low indexing overhead (e.g.~1\% for \texttt{UserVisits} when using an offer rate of 10\%) only for the very first job. The following jobs already run faster than HAIL, e.g.~$\sim\!\!2$ times faster from the fourth job with an offer rate of 25\%. The results also showed that, even for low offer rates, \hail\ converges to a complete index after running only a few number of MapReduce jobs. For example, \hail\ converges to a complete index after 10 jobs with an offer rate of 10\%. All this demonstrates the high efficiency of \hail\ to (i)~balance indexing effort, (ii)~create clustered indexes at job runtime, and (iii)~adapt to users' workloads. The results also showed that the invisible projection technique incurs a negligible overhead (e.g.~$6\%$ on average for \texttt{Synthetic} dataset).

\noindent\textbf{Acknowledgments.} Research partially supported by BMBF. 

\small

\normalsize

\appendix
\section{Lazy Projection}
\label{appendix:lazy_reordering}
In Section~\ref{section:invisible_projection} we presented {\it invisible projection}, a technique for efficiently creating clustered indexes even if incoming MapReduce jobs projects a subset of attributes. The main idea behind invisible projection is to read the unprojected attributes for those data blocks that are proposed by map tasks to the Adaptive Indexer. This way the Adaptive Indexer can align all attributes inside a data block with respect to the indexed attribute. Our results from Section~\ref{section:exp_invisible_projection} shows that the invisible projection incurs a very low overhead on average. However, invisible projection might incur higher overheads when MapReduce jobs project only a low percentage of attributes. For example, we observe in Figure~\ref{figure:invisiblePavlo} that invisible projection incur an overhead of $\sim\!\!45\%$ when projecting only the first attribute (out of eight) from the \texttt{UserVisists} dataset.

We thus propose {\it lazy projection}, a technique that allows \hail\ to efficiently create clustered indexes when MapReduce jobs project only a small proportion of attributes. In contrast to the invisible projection technique, the main idea of lazy projection is to read only the projected attributes in order to minimise the additional read I/O for partial projections. This means that, with the lazy projection technique, \hail\ reads exclusively the attributes requested by users. Thus, map tasks might pass potentially incomplete blocks to the Adaptive Indexer. The Adaptive Indexer, in turn, apply sorting and reordering only on the available subset of attributes. 

Let's consider again our job example of Section~\ref{section:invisible_projection}. Recall that $job_d$ filters records based on attribute~$d$ and projects only attribute~$b$. In this example, using the lazy projection, the Index Builder first sorts attribute~$d$ and thereby create a permutation vector as described in Section~\ref{section:adaptive-indexer_indexbuilder}. Then, the Index Builder reorders the available attribute~$b$ according to the permutation vector. Finally, the Index Writer creates the corresponding metadata and stores the just created partial clustered index as a {\it partially} pseudo data block replica. In contrast to pseudo data block replica, a partially pseudo data block replica contains the permutation vector of attribute~$d$. This permutation vector indicates the Index Builder how to reorder attributes for alignment w.r.t. the indexed attribute~$d$.

Now, assume that another incoming job $job_d'$, with a filter condition on attribute~$d$, which also projects attributes $c$ besides attribute~$b$. In this case, the \hail\ RecordReader uses the previously created partially pseudo data block replicas to perform an index access on attribute~$d$ so as to read only the qualifying values from attribute~$b$. Additionally, the \hail\ RecordReader reads the persisted permutation vector of attribute~$d$. Then, the \hail\ RecordReader uses the normal data block replica (stored locally) to load the missing attribute~$c$. Once attribute~$c$ is main memory, the \hail\ RecordReader uses the permutation vector to pass the qualifying values from $c$, together with those from $b$, to the map function. When all qualifying records are passed to the map function, the \hail\ RecordReader then passes the permutation vector and all the data from attribute~$c$, which is already in main memory, to the Adaptive Indexer. Internally, the Index Builder reorders attribute~$c$ according to the permutation vector and passes the result to the Index Writer. Finally, the Index Writer locates the matching partially pseudo data block replica, appends the aligned attribute~$c$, and updates the block metadata. In other words, lazy projection tolerates missing attributes in partially pseudo data block replicas. Lazy projection incrementally completes a partially pseudo data block replica whenever the replica (i)~has the right index to perform an incoming MapReduce job, and (ii)~does not contain all the attributes projected by the incoming job. Once a partially pseudo data block replica is complete, i.e.~it contains all attributes, the Index Writer simply deletes the permutation vector.

Therefore, the advantage of lazy projection is that it allows \hail\ to fully incrementally adapt its indexes to users' workloads. Furthermore, lazy projection also allows \hail\ to reduce space consumption, because only those attributes that are actually required by MapReduce jobs are stored in pseudo data block replicas.

\end{sloppypar}
\end{document}